\documentclass[iop]{emulateapj}
\usepackage{graphicx}
\usepackage{amsmath}
\usepackage{mathrsfs} 
\usepackage{pstricks,pst-plot}

\newcommand{\ud}{\mathrm{d}}
\newcommand{\pd}{\partial}
\newcommand{\udx}[1]{\frac{\ud}{\ud #1}}
\newcommand{\pdx}[1]{\frac{\pd}{\pd #1}}
\newcommand{\phx}{{\texttt {PHOENIX}}}
\renewcommand{\l}{\lambda}

\begin{document}

\title{Radiation Energy-Balance Method for Calculating the Time Evolution of Type Ia Supernovae During the Post-Explosion Phase}
\author{Daniel R. van Rossum}
\affil{Department of Astronomy and Astrophysics, University of Chicago, Chicago,IL 60637, USA}
\affil{Center for Astrophysical Thermonuclear Flashes, University of Chicago, Chicago,IL 60637, USA}


\begin{abstract}
A new method is presented for calculating the time evolution of spherically symmetric Type Ia Supernova in the post-explosion phase, enabling light curves and spectra to be simulated in a physically self-consistent way.
The commonly exploited radiative equilibrium, that is in essence a \emph{gas energy balance} condition, is unsuitable for this purpose for important physical and numerical reasons.
Firstly, the RE depends on the heating and cooling rates of the gas by the radiation field, two quantities that almost completely cancel and are very difficult to calculate accurately.
Secondly, the internal energy of the gas is only a tiny fraction of the total energy in the system (the vast majority of the energy resides in the radiation field), so that the vast majority of the energy is neglected in solving for the energy balance.
The method presented in this paper, based on the \emph{radiation energy balance}, addresses the bulk of the energy, does not depend on the heating/cooling rates, guarantees an accurate run of the bolometric luminosity over time while bringing the gas temperatures into consistence with the radiation field.
We have implemented the method in the stellar atmosphere code \phx{} and applied it to the classical W7 model.
The results illustrate the importance of each of the four physical contributions to the energy balance as a function of time.
The simulated spectra and light curves for W7 show good resemblance to the observations, which demonstrates what can be done using \phx{} with the REB method.

\end{abstract}

\section{Introduction}
Type Ia Supernovae (SNe Ia) can be divided physically into two phases: the explosion phase and the ballistic or free-expansion phase.

In the first phase, a white dwarf explodes.
During the explosion, the material in the white dwarf undergoes a series of thermonuclear reactions (\cite{Hoyle60}, \cite{Hillebrandt00}).
These reactions release a large amount of energy and produce heavy chemical elements.
The energy that is released heats up the material and accelerates it to high velocities.
The explosion process takes place in seconds.  The resulting velocities are so large that the gravitational forces rapidly drop as the material expands, and the pressure gradients rapidly decrease as the material expands and cools.
Thus the whole structure becomes ballistic, meaning that the velocity structure of the material no longer changes.

In the second phase, another energy release mechanism becomes important.
Instead of the expanding ejecta cooling down and becoming dim, the radioactive decay of unstable isotopes produced in the explosion phase heat it, making it glow (\cite{Truran67}, \cite{Colgate69}).
The heating occurs via the gamma rays that are produced in the decay of these isotopes, which lose part of their energy as they pass through the expanding ejecta.
The main contribution comes from $\phantom{}^{56}$Ni, which decays through unstable $\phantom{}^{56}$Co, to stable $\phantom{}^{56}$Fe.

It is the ballistic phase of SNe Ia that is observable, and it's luminosity is high enough to overpower that of the host galaxy.
The extreme peak luminosities of SNe Ia, as well as their similarities in peak luminosity and spectra, and the correlation that exists between their peak luminosity and the rate they fade make SNe Ia very interesting objects observationally and theoretically.

Successful theoretical accounts of the \emph{qualitative} observational features of SNe Ia date back to the early eighties (explosion phase: \cite{Nomoto84} and ballistic phase: \cite{Branch85}).
Since then, the theory and methods for modeling both phases have evolved, but reproducing the \emph{detailed} observational features of SNe Ia quantitatively is still a challenge.
The impetus to increase the complexity and sophistication of the models will therefore continue for the foreseeable future.

One recent development is the transition from 1D to 2D and 3D explosion models.
In 3D explosion models using state-of-the-art mechanisms (like deflagration-to-detonation transition [see, e.g., \citet{Kasen09}] and gravitationally confined detonation [see e.g., \citet{Jordan08}]), significant deviations from spherical symmetry in composition and density are found.
These angular inhomogeneities require the \emph{radiative transport} models for the ballistic phase also to be 3D, because it is not possible to determine the impact of a 1D approximation (i.e., averaging over solid angle) without first doing the full 3D treatment.

Although 3D radiation transport codes suitable for SN Ia models are available (\cite{Hauschildt10}, \cite{Kasen08}), the computational demands are currently overwhelming.
This limits the amount of detail that can be included in other areas; e.g., resolution in space, time and momentum, and the ability to treat the interactions between the radiation field and matter micro-physically using statistical equilibrium (called non--local thermodynamic equilibrium (NLTE)), rather than using local thermodynamic equilibrium (LTE).
The impact of such approximations on multi-D models can be estimated from 1D models, in which the physics other than the multi dimensionality can be treated in most detail.

In subsequent papers, we will study the effects of NLTE including ionization that results from the non-Maxwellian electron distribution.
In this paper, we present a new method for solving the radiation-hydrodynamical energy equation (with fixed velocities), assuming spherical symmetry, that makes it possible to evolve SNe Ia from the end of the explosion phase throughout the ballistic phase, and therefore to simulate light curves and spectra, in a physically self-consistent way.

\section{Outline of the problem} \label{sec:ProblemOutline}
SNe Ia in their ballistic phase are radiation-hydrodynamical systems.
The hydrodynamical aspect becomes trivial if the velocities are fixed (which is a reasonable approximation\footnote{\cite{Pinto00} and \cite{Woosley07} studied post-explosion acceleration effects and find changes to the velocity structure of up to 10\%.} -- hence the term ``ballistic phase'').
Given this approximation, the evolution of the density structure of the ejecta with time is then known a priori.
Nevertheless, the fact that the material is moving is \emph{crucial} in the treatment of the radiation.
The dynamical equations for a radiating fluid contain velocity-dependent terms of the order $(v/c)$, unlike the situation for non-radiating fluids, where the frame-dependent terms are only $O(v^2/c^2)$ and are negligible even for the high velocities found in SN Ia ejecta, as pointed out by \cite{Castor72}.

Via the ideal gas law, and if we assume LTE, the caloric equation of state (i.e., the Saha-Boltzmann equations), the gas pressure and internal energy are related to the temperature, so that there is only one variable related to the properties of the gas left in the system\footnotemark.
\footnotetext{In NLTE the situation becomes more complicated, where formally tens of thousands of indirectly coupled gas variables (namely, the occupation numbers of all atomic levels) are present in the system.
The method presented in this paper can naturally be applied to NLTE models, but we will postpone a detailed discussion of NLTE until the next paper (which is in preparation).
}
There are no a priori constraints on the radiation field variables.
The equations describing the system are the radiative transfer equation (RTE), the gas energy equation (GEE) and the radiation energy equation (REE).
The mechanical energy equation, which is used to compute changes in the velocity structure, does not need to be considered in the ballistic phase since the velocities do not change.

Although the general RTE explicitly includes temporal variations of the radiation field, this term has been shown to often be unimportant in astrophysical flows (\cite{Castor72,Buchler79,Mihalas84}) and in SNe Ia specifically by \cite{Baron96}.
\cite{Kasen06} found that SN Ia spectra obtained taking into account the full time-dependence do not differ much from time-independent calculations with the correct boundary conditions (inferred from the full calculations).
Accordingly, the time-dependence of the RTE is weak (and neglected in this work), so that at every instant of time the radiation field is in large part determined by the temperature structure at that time.

In this picture, the GEE and REE provide constraints on the temperature, involving the wavelength-integrated angular moments of the radiation field.
While the GEE and REE are time-dependent, the temperature at a single point in space-time is formally coupled, through the radiation field and in a highly non-linear way, to the temperatures at all other points in space and time before the current time.

Therefore, a temperature-correction procedure is needed that converges to a solution of the temperature on the whole space-time grid that satisfies the GEE and REE conditions.

One way to approach this problem was proposed in \cite{Jack09} but it was not applied to a SN Ia problem.
In their most realistic test scenario two of the four important energy terms (the flux gradient and the work done by the radiation field) together referred to as ``structure term'' were dropped (their energy equation 22).
In two follow-up papers (\cite{Jack11,Jack12}) a different approach, based on the GEE, is proposed and applied to a classical SN Ia problem.
But the presented light curves strongly disagree with the observations in that the early magnitudes drop monotonically from the U-band all through the BVRIJHK-bands with 2.5 magnitudes (U-I Band in Figures 12-16 of \cite{Jack11} and I-K bands in Figures 2-5 of \cite{Jack12}), a fact that is obscured in the Jack et al. papers by shifting the observed light curve data to the model.
Furthermore, in \cite{Jack11} it is described that the method is too computationally intensive to calculate the whole light curve and that big forward jumps in time have to be taken in which the time-dependence (their Equation\,(1)) is \emph{not tracked}.
Consequently, the scheme is physically inconsistent and does not lead to accurate light curves and spectra.

\section{Modified radiative equilibrium: the caveats} \label{sec:ModifiedRE}
In the ballistic phase, there is only one unknown gas variable and that is the internal energy (or equivalently gas temperature or gas pressure), as explained above.
Considering all mechanisms of energy transfer to and from the material (i.e., the first law of thermodynamics) leads to the gas energy equation.
The GEE in the co-moving frame, accurate to $O(v/c)$, is given by (\cite{Mihalas84}, Equation (96.7))
\begin{equation} \label{eq:GEEorig}
 \rho \left[ \udx{t} e + p \udx{t} \frac{1}{\rho} \right]
  = 4\pi Q + \rho \varepsilon \; .
\end{equation}
Here $\rho$ is the density, $e$ is the internal energy per unit mass, $p$ is the gas pressure, $\varepsilon$ is the heating of the gas by gamma rays\footnotemark{} per unit mass per unit time, and $4\pi Q$ is the net heating by radiation per unit volume per unit time, where
\footnotetext{The heating due to positron annihilation is included, and the gamma photons produced are transported like primary gamma photons.}
\begin{equation} \label{eq:GasHeating}
 Q \equiv \int_0^\infty \chi_\l J_\l - \eta_\l \,\ud\l \; ,
\end{equation}
$\chi_\l$ and $\eta_\l$ are the radiative monochromatic extinction coefficient and emissivity, $J_\l$ is the monochromatic mean intensity, and $\l$ is the wavelength.
Equation \eqref{eq:GEEorig} states that the rate of change of the material energy density plus the rate of work done by the material pressure equals the net rate of energy input from the radiation field and thermonuclear sources.

\begin{figure}
 \centerline{\includegraphics[width=.47\textwidth]{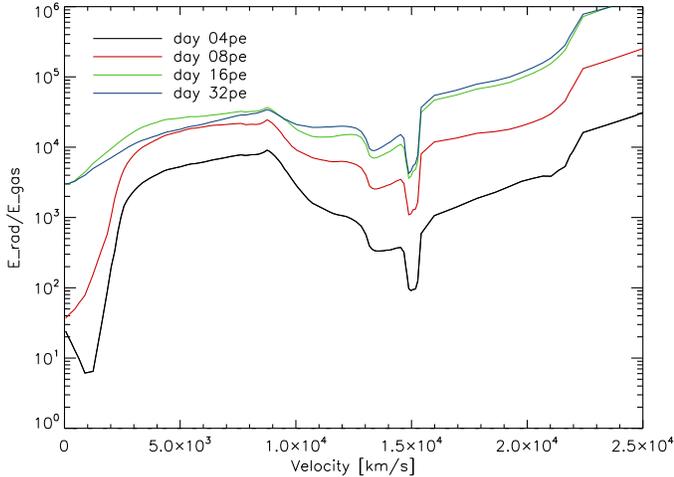}}
 \caption{The ratio of radiation energy density $E_{\rm rad}$ over gas energy density $E_{\rm gas}$ is very high throughout a SN Ia structure at different times after the explosion (here the results for the W7 model are shown).
 $E_{\rm gas} \propto \rho T$ does not gain as much from the high temperatures at low densities as the radiation energy density, which roughly scales like $E_{\rm rad} \propto T^4$ (see e.g. \cite{Mihalas84}).\\
 The low ratio values at low velocities reflect the absence of radioactive material around the center in the W7 model (mass fraction $< 10^{-7}$), which at early times is too opaque for gamma rays to penetrate and thus remains cool.} \label{fig:Edensity}
\end{figure}

Evaluating the terms of Equation \eqref{eq:GEEorig} in SN Ia models shows that after a few days from the explosion (depending on the abundance of radioactive material throughout the structure) the terms on the left-hand side are negligibly small compared to the two terms on the right-hand side.
The reason is that the internal energy density of the material is extremely small compared to the energy density of the radiation field, given the low densities and high temperatures in the ejecta, as shown in Figure \ref{fig:Edensity}.
Thus to a good approximation (after the first few days), the GEE \eqref{eq:GEEorig} reduces to
\begin{equation} \label{eq:GEESimple}
  4\pi Q = - \rho \varepsilon \; ,
\end{equation}
which is similar to but is not the strict radiative equilibrium (RE) condition $Q = 0$.

Equation \eqref{eq:GEESimple} offers a tempting, simple way to determine the temperature structure of the ejecta.
The net heating term $Q$ can be evaluated by doing the full radiation transport.
Departures from the condition expressed by Equation \eqref{eq:GEESimple} can be corrected for in a quasi-local fashion, since in zeroth order approximation, locally the net heating by the radiation field increases with decreasing local temperature.
Corrections for non-local effects (e.g., the fact that the local net radiative heating effect decreases with decreasing global temperatures) can be accounted for iteratively.

There are three major problems with this formalism.
Firstly, the opacity changes dramatically over small wavelength scales (i.e., a line width), so that the wavelength sampling of the opacity becomes important in evaluating the integral in Equation \eqref{eq:GasHeating}.
Also, the result changes if lines are omitted.
Secondly, because $Q$ is a small difference between two large numbers [see Equation \eqref{eq:GasHeating}] -- it is typically smaller by a factor of $10^3 - 10^7$ than the sum of both -- the precise value of $J$ near line centers becomes important, since there the weighting is the strongest.
Therefore, this method is very sensitive to approximations in the radiative source function or inaccuracies in the solution of the RTE.
Finally, the GEE is not the only energy balance condition.
The vast majority of the energy is stored in the radiation field, and this method does not take into account the local battery effect of the storage of energy in, and at later time the retrieval of energy from, the radiation field (see section \ref{sec:Results}).

Indeed, the gas can not absorb a significant amount of the energy deposited by radioactive decay, because through material-radiation interactions it is directly transferred to the radiation field, with little effect on the temperature.
Let us therefore now turn our attention from the material to the radiation, and consider the physical effects on the energy density of the radiation field.

\section{The Radiation Energy Balance (REB) method} \label{sec:Method}
\subsection{The target flux}
The REE in the co-moving frame in spherical symmetry, accurate to $O(v/c)$, is given by (\cite{Mihalas84}, Equation (96.8))
\begin{multline} \label{eq:REEorig}
 \frac{\rho}{c} \left[ \udx{t} \frac{J}{\rho}
   + K \udx{t} \frac{1}{\rho}
   - (3K - J) \frac{v}{\rho r} \right] \\
   = -Q - \frac{1}{r^2} \pdx{r} (r^2 H) \; ,
\end{multline}
where and $J$, $H$ and $K$ are the zeroth, first, and second angular moments of the radiation field, which physically represent the radiation energy density, the radial flux, and the radiation pressure (see e.g. \cite{Mihalas84}).
$Q$ can now (equivalently) be interpreted as the net \emph{cooling} of the radiation field by the material.
Equation \eqref{eq:REEorig} states that the rate of change of the radiation energy density plus the rate of work done by radiation pressure (the second and third terms on the left-hand side) equals the net rate of energy flowing into the radiation field from the material, minus the net rate of radiant energy flowing out of a fluid element by transport, all per unit mass.

With the high expansion velocities typical of SN Ia, the radial coordinate $r$ of a co-moving grid point quickly (within minutes or less) becomes proportional to its velocity
\begin{equation} \label{eq:Homology}
 r = v t \; ,
\end{equation}
where $t$ is the time since explosion.  The derivative of the inverse density can then be written as
\begin{equation} \label{eq:RadWork}
 \udx{t} \frac{1}{\rho} = \frac{3}{t} \frac{1}{\rho} \; ,
\end{equation}
and the terms proportional to $K$ on the left-hand side of Equation \eqref{eq:REEorig} cancel out.

The next step is to express $Q$ in the REE \eqref{eq:REEorig} in terms of $\varepsilon$ using the GEE \eqref{eq:GEEorig}.
The terms on the left-hand side of the GEE are much smaller than the corresponding terms in the REE (the left-hand side) in the SN Ia problem (see figure \ref{fig:Edensity}) and can be dropped\footnote{Note that dropping the gas energy density terms is not at all required for the REB method, but here supports focusing on the big energy pool, which is the radiation field.}.
The net cooling by the material now does not have to be evaluated (with all of the uncertainties in doing so we mentioned before) because the material does not (significantly) absorb energy and therefore all of the energy flowing into the material flows out to the radiation field.

Using Equations \eqref{eq:GEESimple} and \eqref{eq:RadWork} and the spherical symmetry boundary condition $H(r\!=\!0) = 0$, we can integrate Equation \eqref{eq:REEorig} to obtain a target flux $H_{\rm tg}(r)$ for all radial coordinates $r$ on the grid,
\begin{equation} \label{eq:TargetFlux}
 r^2 H_{\rm tg}(r) = \int_0^r -\frac{\rho r^2}{c}\udx{t} \frac{J}{\rho}
   - \frac{r^2}{ct} J + \frac{r^2 \rho}{4\pi} \varepsilon \,\ud r \; .
\end{equation}
Through this \emph{target flux equation}, the first angular moment of the instantaneous local radiation field is related to the zeroth angular moment and its time derivative of the local and underlying shells.
This is a highly indirect and non-local condition for the instantaneous local temperature, the more so since in a non-grey atmosphere the influence of the temperature on the radiation field involves a very large number of individual opacity sources.

\begin{figure}
 \centerline{\includegraphics[width=.47\textwidth]{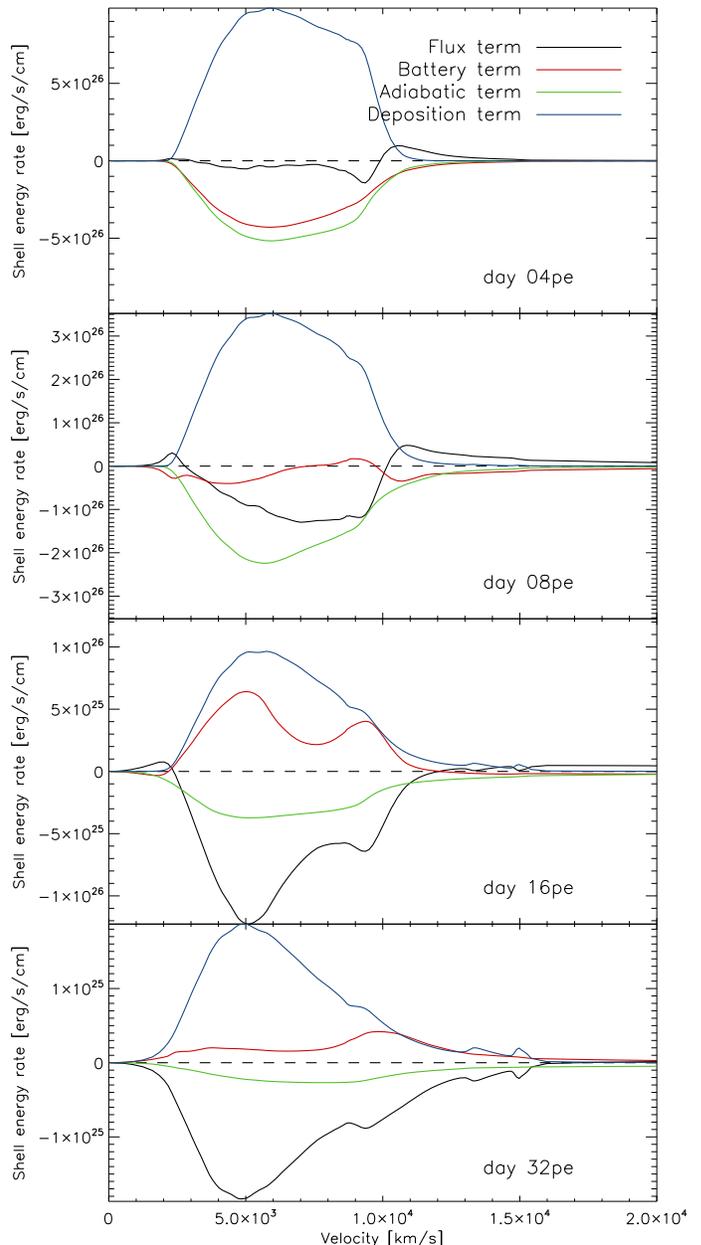}}
 \caption{
 This figure shows the contributions to the rates of energy change of the four terms of the radiation energy equation \eqref{eq:REEorig} (``Battery'' (red) is the first and ``Adiabatic'' (green) the sum of the second and third terms on the left-hand side, while ``Deposition'' (blue) is the first and ``Flux'' (black) the second term on the right-hand side) multiplied by $r^2$, plotted against the velocity for the W7 model at 4, 8, 16, and 32 days after explosion (pe).
 These four terms correspond to the terms in the target flux equation \eqref{eq:TargetFlux} after taking the spatial derivative $\pd/\pd r$.
 The x-axis is limited to $2 \cdot 10^4$ km/s, beyond that all terms are small.\\
 Note that the absolute rates decline rapidly with time.\\
 At early times, energy is stored in the radiation field, and the battery term is negative.
 At day 4pe, the rate of storage of energy in the radiation field is about as high as the work term (``Adiabatic'') caused by the expansion of the co-moving frame (in the observer frame, work would be done on material moving through a radiation pressure gradient).
 These two effects together almost completely balance the heating (``Deposition'') of the radiation field by the gas from the energy deposited in the gas by gamma rays.
 As a result, the gradient of the radial flux (``Flux'') becomes very small.\\
 At later times, the battery term and the term become smaller due to cooling by expansion but continue to be important.
 The relative sizes of the terms are governed by the evolution of the temperature structure as a function of time.
 Methods that neglect these effects in the determination of the temperature structure miss important physics.
 } \label{fig:HtgLayer}
\end{figure}

\begin{figure}
 \centerline{\includegraphics[width=.5\textwidth]{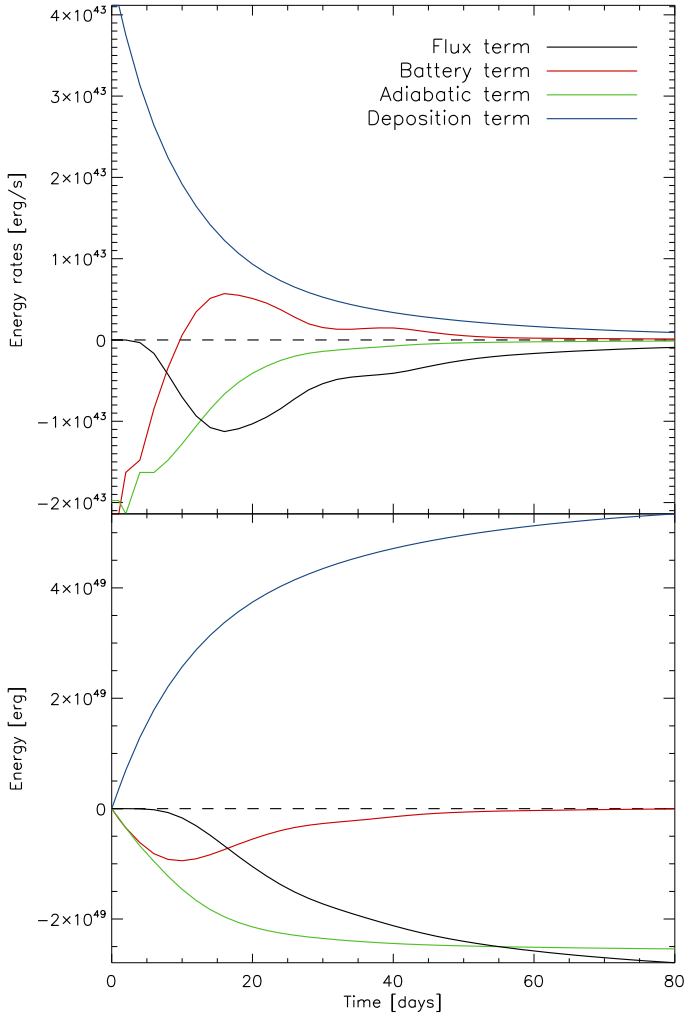}}
 \caption{
 The upper panel shows the contributions of the four terms in the target flux equation \eqref{eq:TargetFlux} for the W7 model: the flux (black), the battery term (red), the adiabatic cooling of the radiation field (green), and the heating of the radiation field by gamma ray absorption (blue) respectively.
 See Figure \ref{fig:HtgLayer} for the radial distribution of these terms.
 Integration of the flux term (Figure \ref{fig:HtgLayer}) from the center outwards yields the total flux $R^2H(R)$ at the outer boundary $R$, which is the luminosity of the structure.
 After about 55 days, the energy flowing into the ejecta from nuclear decays has gone into work and luminosity in equal amounts.\\
 Note that the bolometric luminosity (black curve) is a \emph{direct product} of the REB method.\\
 The lower panel shows the integration over time of these same total rates as a function of time (the quantities are assumed to be constant before t=1 for simplicity).
 It shows that the radiation field battery term integrates to zero.
 This is an important verification of the radiation energy balance (REB) method.
 Physically, the battery term cannot release more energy than what has been stored previously, but this requirement is not directly accounted for, since the target flux equation is in differential form in $t$ and not in integral form.
 Whereas instantaneous energy conservation (the requirement that the sum of the four rate terms be zero) is explicitly solved for in each point in time, this result shows that the $\ud (J/\rho)/\ud t$ computation is accurate and satisfies temporal energy conservation in this time-dependent term, individually.
 } \label{fig:Ebal}
\end{figure}

An example of the contribution of each of the terms of the REE in different regions of the atmosphere and for different evolution times after the explosion is given in Figure \ref{fig:HtgLayer}, where the W7 model \cite{Nomoto84} was used for the explosion phase (see section \ref{sec:TestSetup} for details on the test setup).
These results were obtained by solving the target flux equation using the method described in the next section.
Figure \ref{fig:Ebal} shows the evolution of the different terms in the REE, integrated over the whole structure, as a function of time.

The target flux Equation \eqref{eq:TargetFlux} does not take into account the thermodynamic state at the time of the transition from the explosion phase to the ballistic phase.
In particular, the internal energy $e$ of the gas at the time of transition determines the temperature of the structure before the local heating by gamma rays and/or the ``thermal'' radiation field takes over.
In regions with abundant $\phantom{}^{56}$Ni, this happens only a few minutes after explosion, because the cooling rate due to adiabatic expansion of the gas (which is proportional to $e$) is still very high.
But in the outer shells that have no radioactive Nickel, it takes a few days before the ejecta are optically thin enough for these shells to be heated by gamma rays and thermal radiation originating in deeper shells.
For such conditions (optically very thick outer shells at early times), the target flux Equation \eqref{eq:TargetFlux} cannot be expected to yield realistic results.
However, this is not a problem for the radiation energy balance (REB) method because the contribution of these regions to the global energy balance is insignificant.

The important point is that the energy released by early nuclear decays cannot escape from the ejecta right away; it is stored locally in the radiation field and released later on.
Thus the radiation field acts as a battery; this has an important effect on the \emph{temperature structure} at later times, and therefore on the light curves and spectra.

\subsection{Translation to temperature corrections} \label{sec:TempCorrection}
We do not presume to determine analytically the exact temperatures (or temperature corrections) that satisfy the target flux equation.
Instead, we propose a method that translates departures of the actual flux from the target flux to \emph{approximate} temperature corrections and iterate for convergence to the physically correct solution.
The derivation partly follows \cite{Hauschildt03} and \cite{Lentz03} where the Uns\"old-Lucy method (\cite{Unsoeld55}, \cite{Lucy64}) is generalized to spherical geometry.

We start from the observation mentioned before that the heating and cooling rates of the gas by the radiation are much bigger in size than their sum
\begin{equation} \label{eq:HeatCoolCancel}
 Q^{\rm heat} \approx -Q^{\rm cool} \gg Q^{\rm heat} + Q^{\rm cool} = Q \;,
\end{equation}
where $Q^{\rm heat}$ and $Q^{\rm cool}$ represent the first and second term on the right-hand side of Equation \eqref{eq:GasHeating}
This condition does \emph{not} contain enough physics to determine the dynamics of the system (it contains even less physics than Equation \eqref{eq:GEESimple}), but is adequate for the purpose of driving the model towards the physically justified condition of Equation \eqref{eq:TargetFlux}.

The extinction coefficient and emissivity are assumed to consist of true absorption ($\kappa$) and pure scattering ($\sigma$) contributions $\chi_\l = \kappa_\l + \sigma_\l$, and thermal emission and scattering contributions $\eta_\l = \kappa_\l B_\l + \sigma_\l J_\l$, respectively, where $B_\l$ is the Planck function.
Note that this macro-physical description is suitable for LTE but needs to be generalized for NLTE (see our paper on NLTE in preparation).
From Equations \eqref{eq:GasHeating}, \eqref{eq:GEESimple}, and \eqref{eq:HeatCoolCancel} follows
\begin{equation} \label{eq:RE}
 \int_0^\infty \kappa_\l B_\l \,\ud \l = \int_0^\infty \kappa_\l J_\l \,\ud \l \; ,
\end{equation}
where the scattering contributions cancel.
Let $B$ be the wavelength-integrated Planck function.
If one subtracts Equation \eqref{eq:RE} for the current temperature iteration from the next iteration (marked with primes) and assumes that the wavelength-averaged opacities,
\begin{align*}
 \kappa_B = \frac{1}{B} \int_0^\infty \kappa_\l B_\l \,\ud \l \; , \\
 \kappa_J = \frac{1}{J} \int_0^\infty \kappa_\l J_\l \,\ud \l \; , \\
 \chi_H = \frac{1}{H} \int_0^\infty \chi_\l H_\l \,\ud \l \; ,
\end{align*}
are insensitive to temperature changes (which is a good approximation, as observed by \cite{Lucy64}), Equation \eqref{eq:RE} can be written as
\begin{equation} \label{eq:DeltaB}
 \Delta B = \frac{\kappa_J}{\kappa_B} \Delta J \; ,
\end{equation}
where $\Delta B = B' - B$ and $\Delta J = J' - J$.
The temperature correction $\Delta T$ follows from $\Delta B$ using the temperature derivative of the wavelength integrated Planck law,
\begin{equation} \label{eq:DeltaT}
 \Delta T = \frac{1}{\ud B/\ud T} \Delta B
             = \frac{T}{4} \frac{\Delta B}{B} \; .
\end{equation}

Following \cite{Unsoeld55}, we derive $\Delta J$ from the radiation momentum equation.  In the co-moving frame and assuming spherical symmetry, this equation, accurate to $O(v/c)$, is given by (\cite{Mihalas84}, Equation (95.21))
\begin{multline} \label{eq:RadMomentum}
 \pdx{r} K + \frac{3K - J}{r} 
   = -\chi_{H} H - \frac{2}{c} \left( \frac{\partial v}{\partial r} + \frac{v}{r} \right) H \\
     - \frac{1}{c} \udx{t} H - \frac{1}{c} \frac{\ud v}{\ud t} \left( J + K \right) \; .
\end{multline}
Using the \emph{sphericity factor} $q$ introduced by \cite{Auer71},
\begin{equation*}
 q \equiv \frac{r_c^2}{r^2} \exp \left[ \int_{r_c}^r \frac{3f-1}{fr'} \ud r' \right] \; ,
\end{equation*}
where $f=K/J$ is the Eddington factor and $r_c$ is a reference radius, the left-hand side of Equation \eqref{eq:RadMomentum} can be rewritten, and the right-hand side can be evaluated for Equation \eqref{eq:Homology}, yielding,
\begin{equation} \label{eq:RadMomentumHomology}
 \frac{1}{q r^2} \pdx{r} (qf r^2 J) = - \left( \chi_H + \frac{4}{ct} \right) H - \frac{1}{c} \udx{t} H \; .
\end{equation}
Subtracting Equation \eqref{eq:RadMomentumHomology} for the current temperature iteration from the next iteration and assuming that $f$ is insensitive to the temperature\footnote{
Although this is typically a good approximation, it does not have to hold exactly, given the intentional approximate nature of the correction being calculated.}
one obtains,
\begin{equation} \label{eq:RadMomentumDelta}
 \frac{1}{q r^2} \pdx{r} (qf r^2 \Delta J) = - \left( \chi_H + \frac{4}{ct} \right) \Delta H \; .
\end{equation}
Here $\Delta H = H' - H = H_{\rm tg} - H$.
The time derivative in Equation \eqref{eq:RadMomentumHomology}, evaluated on a discrete time grid against the reference time $t^*$, becomes
\begin{equation}
 \udx{t} H = \frac{H_{\rm tg} - H^*}{t - t^*} - \frac{H - H^*}{t - t^*} = \frac{\Delta H}{t - t^*} \; .
\end{equation}
The numerator on the right-hand side is independent of the size of the time step in the denominator so that this term has to be dropped.
This is the case because the reference value $H^*$ is the same for both temperature iterations.
Integration of Equation \eqref{eq:RadMomentumDelta} from the outermost model radius $R$ to $r$ yields
\begin{multline} \label{eq:DeltaJ}
 \Delta J(r) = \frac{1}{q(r) f(r)} \Bigg{(} q(R) f(R) g(R) \,\Delta H(R) \\
  - \frac{1}{r^2} \int_{R}^r q(r') \left( \chi_H(r') + \frac{4}{ct} \right) r^2 \Delta H(r') \,\ud r' \Bigg{)} \; ,
\end{multline}
where $g = J/H$ is the second Eddington factor, which is assumed to be insensitive to the temperature for the outermost model layer at radius $R$.  From Equations \eqref{eq:DeltaB}, \eqref{eq:DeltaT} and \eqref{eq:DeltaJ}, we obtain the final expression for the temperature correction:
\begin{multline} \label{eq:TempCorrection}
 \Delta T = \frac{T}{4B} \frac{\kappa_J}{\kappa_B} \frac{1}{q(r) f(r)} \Bigg{(} q(R) f(R) g(R) \,\Delta H(R) \\
  - \frac{1}{r^2} \int_{R}^r q(r') \left( \chi_H(r') + \frac{4}{ct} \right) r^2 \Delta H(r') \,\ud r' \Bigg{)} \; .
\end{multline}
All of the quantities on the right-hand side of Equation \eqref{eq:TempCorrection} are available upon completion of each temperature correction iteration.
Note that both $H$ and $H_{\rm tg}$ change with temperature, and iteration is required to converge $H$ to $H_{\rm tg}$.

The number of iterations required is found to range between 3 and 10, using the Active Damping method to accelerate convergence (paper in preparation), depending on the assumed initial temperature structure and the relative contributions of the terms in the target flux equation (the constant third term is the fastest to converge and the time-dependent first term is the slowest).

\subsection{Sequential solver for the time-dependence} \label{sec:TimeGrid}
With the target flux equation (Equation \eqref{eq:TargetFlux}) and the temperature correction equation (Equation \eqref{eq:TempCorrection}), the formalism and the tools are available that are needed to solve, in principle, the radiation-hydrodynamical evolution of a SNe Ia during the ballistic phase.

Two difficulties arise in the numerical evaluation of the time-dependence in the target flux equation (Equation \eqref{eq:TargetFlux}).
Firstly, the solution at location $r$ and time $t$ depends formally on the history of the whole structure, so that we face an initial condition problem.
The initial condition for the time-dependence is
\begin{equation} \label{eq:J0BoundaryCondition}
 \left. \frac{J}{\rho} \right|_{t\!=0} = 0 \; ,
\end{equation}
which physically does not hold exactly, but is true to a very good approximation.

\begin{figure}
 \centerline{\includegraphics[width=.5\textwidth]{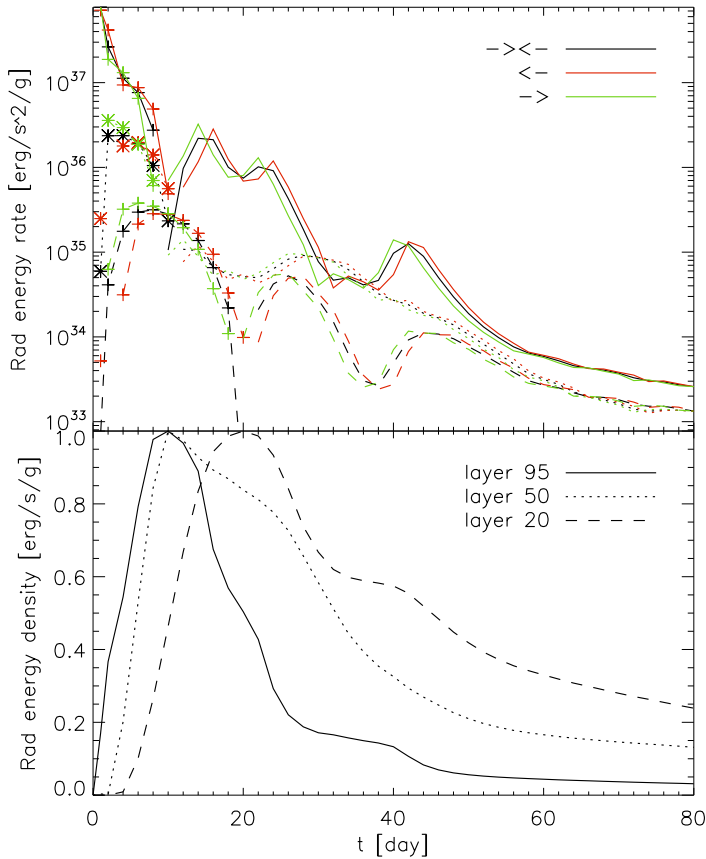}}
 \caption{This figure shows the run of the radiation energy density per mass over time for three different layers in the atmosphere on a normalized linear scale (lower panel).
 These curves have been computed with the model setup described in sections \ref{sec:TestSetup} and \ref{sec:Results}.
 Layer 95 is a rather deep, $\protect\phantom{}^{56}$Ni rich layer, layer 20 lays further outside, far from $\protect\phantom{}^{56}$Ni rich regions. Layer 50 does not contain $\protect\phantom{}^{56}$Ni but lies close to regions that do.
 In early times the structure is optically thick for gamma rays so that layers further out are not yet heated by the nuclear decay.
 Therefore, the rise in energy density sets in later in the outer layers 20 and 50 than in the deeper layer 95.\\
 The slopes of these curves, determined using the 'both-sided' ($\to\!\gets$), and the two 'single-sided' ($\to$ and $\gets$) approximations, are plotted in the upper panel on a logarithmic scale, where positive slopes are indicated with symbols and negative slopes without symbols.
 The single-sided approximations tend to over- or underestimate the slope in non-linear situations. 
 } \label{fig:JSlope}
\end{figure}

Secondly, on a discrete time grid the time derivative can only be approximated, yet accuracy is important.
Let $x=J/\rho$; then the ``both-sided'' linear approximation is
\begin{equation} \label{eq:BothSidesDerivative}
 \udx{t} x_i = \frac{x_{i+1} - x_{i-1}}{t_{i+1} - t_{i-1}} \; ,
\end{equation}
and the two ``single-sided'' linear approximations are
\begin{align}
 \udx{t} x_i &= \frac{x_i - x_{i-1}}{t_i - t_{i-1}} \label{eq:FromLeftDerivative} \; , \\
 \udx{t} x_i &= \frac{x_{i+1} - x_i}{t_{i+1} - t_i} \label{eq:FromRightDerivative} \; .
\end{align}
In the following, we refer to these three variants symbolically as $\to\!\gets$, $\to$ and $\gets$, respectively.
If the variation in $J$ is non-linear, the both-sided approximation clearly is more accurate than the single-sided approximations.
This is demonstrated in Figure \ref{fig:JSlope}.
where both $J/\rho$ and $\ud (J/\rho)/\ud t$ are shown for the model described in sections \ref{sec:TestSetup} and \ref{sec:Results}.
If $J$ progressively increases, $\to$ under(over)-estimates and $\gets$ over(under)-estimates the \emph{size} of the positive slope, whereas if  $J$ progressively decreases, $\to$ under(over)-estimates and $\gets$ over(under)-estimates the size of the negative slope.

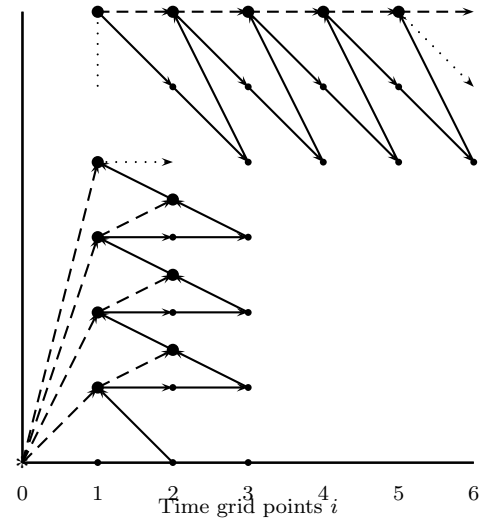
\begin{figure}
\begin{center}
 \psset{arrows=->}
\begin{pspicture}(0,-.4)(6,6)
 \psset{dotsize=1pt 2}
 \uput*{0}[0](1.8,-.6){Time grid points $i$}
 \psaxes[ticks=none,axesstyle=none,labels=x](6,6)
 \psdots*[dotsize=3pt 2](1,4) 
 \psdots*[dotsize=3pt 2](2,3.5)
 \psdots*[dotsize=3pt 2](1,3) \psdots*(2,3)(3,3)
 \psdots*[dotsize=3pt 2](2,2.5)
 \psdots*[dotsize=3pt 2](1,2) \psdots*(2,2)(3,2)
 \psdots*[dotsize=3pt 2](2,1.5)
 \psdots*[dotsize=3pt 2](1,1) \psdots*(2,1)(3,1)
 \psdots*(1,0)(2,0)(3,0)
 \psdots[dotstyle=asterisk, dotsize=3pt 2](0,0)
 \psline[linestyle=dotted](1,4)(2,4) 
 \psline[linestyle=dashed](0,0)(1,4) \psline(2,3.5)(1,4)
 \psline(1,3)(2,3) \psline(2,3)(3,3) \psline[linestyle=dashed](1,3)(2,3.5) \psline(3,3)(2,3.5)
 \psline[linestyle=dashed](0,0)(1,3) \psline(2,2.5)(1,3)
 \psline(1,2)(2,2) \psline(2,2)(3,2) \psline[linestyle=dashed](1,2)(2,2.5) \psline(3,2)(2,2.5)
 \psline[linestyle=dashed](0,0)(1,2) \psline(2,1.5)(1,2)
 \psline(1,1)(2,1) \psline(2,1)(3,1) \psline[linestyle=dashed](1,1)(2,1.5) \psline(3,1)(2,1.5)
 \psline[linestyle=dashed](0,0)(1,1) \psline(2,0  )(1,1)
 \psline[linestyle=dotted,arrows=-](1,5)(1,6)
 \psdots*[dotsize=3pt 2](1,6)(2,6)(3,6)(4,6)(5,6)
 \psdots*(     2,5)(3,5)(4,5)(5,5)
 \psdots*(          3,4)(4,4)(5,4)(6,4)
 \psline(1,6)(2,5) \psline(2,5)(3,4) \psline(3,4)(2,6) \psline[linestyle=dashed](1,6)(2,6)
 \psline(2,6)(3,5) \psline(3,5)(4,4) \psline(4,4)(3,6) \psline[linestyle=dashed](2,6)(3,6)
 \psline(3,6)(4,5) \psline(4,5)(5,4) \psline(5,4)(4,6) \psline[linestyle=dashed](3,6)(4,6)
 \psline(4,6)(5,5) \psline(5,5)(6,4) \psline(6,4)(5,6) \psline[linestyle=dashed](4,6)(5,6)
 \psline[linestyle=dotted](5,6)(6,5)                   \psline[linestyle=dashed](5,6)(6,6)
\end{pspicture}
\end{center}
\caption{
This sequential scheme, consisting of two stages, describes how the time-dependence is solved in the radiation energy balance method: (1) startup conditions (lower left part) and (2) the evolution of the startup conditions to later times (upper right part).
 The horizontal axis shows the time grid points $i$.
 Every dot in the graph represents a model step where the time derivative is ``zero'', ``single-sided'', or ``both-sided'', which is indicated by the number of arrows pointing \emph{in}, being zero, one, or two, respectively. The both-sided steps are indicated with thicker points.
 The star at $i=0$ is the fixed initial condition.
 The solid line shows the sequence of model steps that need to be computed, while the dashed lines indicate the additional time-dependences.
 The dotted arrows indicate that the scheme can be trivially extended in that direction.
 The dotted vertical line in the upper left corner means that  the result provided by the lower left part is used for the first point of the upper right part of the scheme.
} \label{fig:Scheme}
\end{figure}

We propose a sequential scheme to solve the time-dependence that makes use of the both-sided derivatives.
The sequence of steps taken in the scheme are shown in the graph of Figure \ref{fig:Scheme}.
There are two stages.

In the first stage the proper startup conditions are determined, and in the second stage they are evolved to later times.
The time derivative is set to zero initially and the temperature structure is computed individually for the time grid points.
The problem of determining the proper startup conditions is equivalent to accurately determining the time derivative at early times after the explosion.
As it turns out, the variation of $J$ with time is highly non-linear at early times, so that relying on $\to$ is systematically inaccurate.
In order to reduce this systematic effect, $\to$ is replaced by the sequence $\to, \to, \to\!\gets$, where $\to\!\gets$ acts on the starting and end grid points of the two $\to$ operations, $i=1$ and $i=3$ respectively, yielding an improved value at point $i=2$.
In each of these three steps, the temperature structure is iterated to convergence.
When the $\to$ operation is used, the derivative is updated ``on the fly'' with the results of the updated temperature structure, Equation \eqref{eq:FromLeftDerivative}.
With this result in hand, the derivative at $i=1$ is updated with a $\to\!\gets$ operation using the boundary condition \eqref{eq:J0BoundaryCondition}.
Whereas initially the temperature structure for $i=1$ was derived assuming a time derivative of zero, equivalent to assuming a constant $J/\rho$ over time, these steps need to be iterated until the derivative converges for $i=1$.
This typically takes about 4 cycles.

In the second stage, the converged startup condition is evolved in time.
Just as in the first stage, every $\to$ operation is replaced by the sequence $\to, \to, \to\!\gets$, so that every point is computed three times.
The second and third time a point on the time grid is computed, the temperature converges quickly, because the corrections are relatively small (i.e., smaller than the changes from one to the next time step as in the first time).

Note that the use of $\gets$ has been avoided in the scheme.
The reason is that when the $\gets$ operation is used, the derivative can not be updated on the fly with the results of the updated temperature structure, because the dependence of the target flux on the temperature is now inverted, which inevitably leads to poor convergence or even divergence of the temperature-correction method.
Therefore, it would take much longer to achieve convergence using $\gets$ operations.

An alternative way to solve the time-dependence would be to solve the problem on the whole time grid simultaneously.
Again, the time derivative is set to zero initially, so that the temperature structure can be determined for every point in time independently.
Then at every point on the space-time grid, the time derivative is computed using $\to\!\gets$ and the temperature structures are updated simultaneously.
This method is relatively computationally intensive, because in every temperature-correction iteration the radiation transport must be solved on the whole time (and space and momentum) grid, but it parallelizes perfectly.
Therefore, the actual model runtime would not increase if the number of processors is scaled with the number of time grid points.
Even though this method is less efficient, since it doesn't directly exploit the causal property of the evolution, it may be more accurate, because the use of single-sided steps can be completely avoided.
Implementation of this alternative method has not been completed at this time, so we cannot give concrete results about its performance or benefits in accuracy.

\section{Implementation and test model assumptions} \label{sec:TestSetup}
We implemented the formalism and solution scheme described in the previous section in \phx{}, a state-of-the-art, general purpose stellar atmosphere code (\cite{Hauschildt92}, \cite{Hauschildt99}, \cite{Hauschildt04}, \cite{Hauschildt06}).
Wide applicability of the code to astrophysical objects ranging from brown dwarfs to main sequence stars, AGB stars, novae and Supernovae, means the code has been well tested for a broad range of physical conditions.
Specifically, a lot of work on SNe Ia has been done with \phx{}, concentrating mainly on computing snapshots in time (e.g. \cite{Nugent97}, \cite{Lentz01}, \cite{Baron06}) for explosion models like W7 \cite{Nomoto84}.
\phx{} solves the fully special relativistic radiation transport equation on a co-moving grid, including gamma-ray transport (see \cite{Lentz01} and references therein), where a standard value for the effective absorptive gamma-ray opacity of $\kappa_\gamma = 0.06 \,Y_e$ cm$^2$ g$^{-1}$ is assumed.
Here $Y_e$ is the total electron number density divided by the baryon density.
This method is less accurate than a full Monte Carlo treatment, but deviations are small as shown by \cite{Swartz95}.

For the purpose of testing the REB method and qualitatively investigating the effects of the different terms in the energy balance equation, we use the well known W7 explosion model as presented in \cite{Branch85}.
This model has become an established 1D reference model that is known to reproduce qualitatively the observed spectral properties of SNe Ia.
Furthermore, we assume LTE with a line absorption probability $\varepsilon = 0.3$, based on \cite{Kasen06}, and defer the study of NLTE effects to a subsequent paper (in preparation).
The ejecta are assumed to be in the ballistic phase, so velocities are constant in time.
As mentioned before, we neglect the influence of the internal energy of the gas at the time of the transition from the explosion phase to the ballistic phase on the temperature in the early stages of the ballistic phase.

We follow the evolution with time steps of 2 days, starting at $t=4$ days.
With time steps that are too big, the discrete approximations to the time derivatives (Equations \eqref{eq:BothSidesDerivative} to \eqref{eq:FromRightDerivative}) become inaccurate, while with time steps that are too small, the approximations become unstable (because the denominator becomes small).
A time step of 2 days has been found to be a good compromise, but this choice may be optimized (e.g., allowed to vary with time) in the future to improve accuracy and performance.

The number of points computed in the scheme (see Figure \ref{fig:Scheme}) is: 3 without time dependence, 17 in stage 1, and 69 in stage 2.
The computation time in minutes per point in the scheme on 16 Opteron (2.2GHz) processors is: 30 for the first 3 points, 15 for the next 17 points, and 15 or 10 for the last 69 points (15 for the first time, 10 for the second and third times a time grid point $i$ is computed).
The total time for calculating the light curves and spectra is thus $3\cdot 30 + 17\cdot 15 + 22\cdot 35 + 30) = 1145$ minutes, or 19 hours.
Scaling beyond 16 processors is not very good\footnote{For NLTE models the situation is different.  These were found to scale well up to the number of layers used in the model, typically 128. This is important, since NLTE models are computationally much more demanding.}.
Using 32 processors reduces the total time by 15\%.

\section{Results} \label{sec:Results}
\subsection{Light curves}
\begin{figure*}
 \centerline{\includegraphics[width=\textwidth]{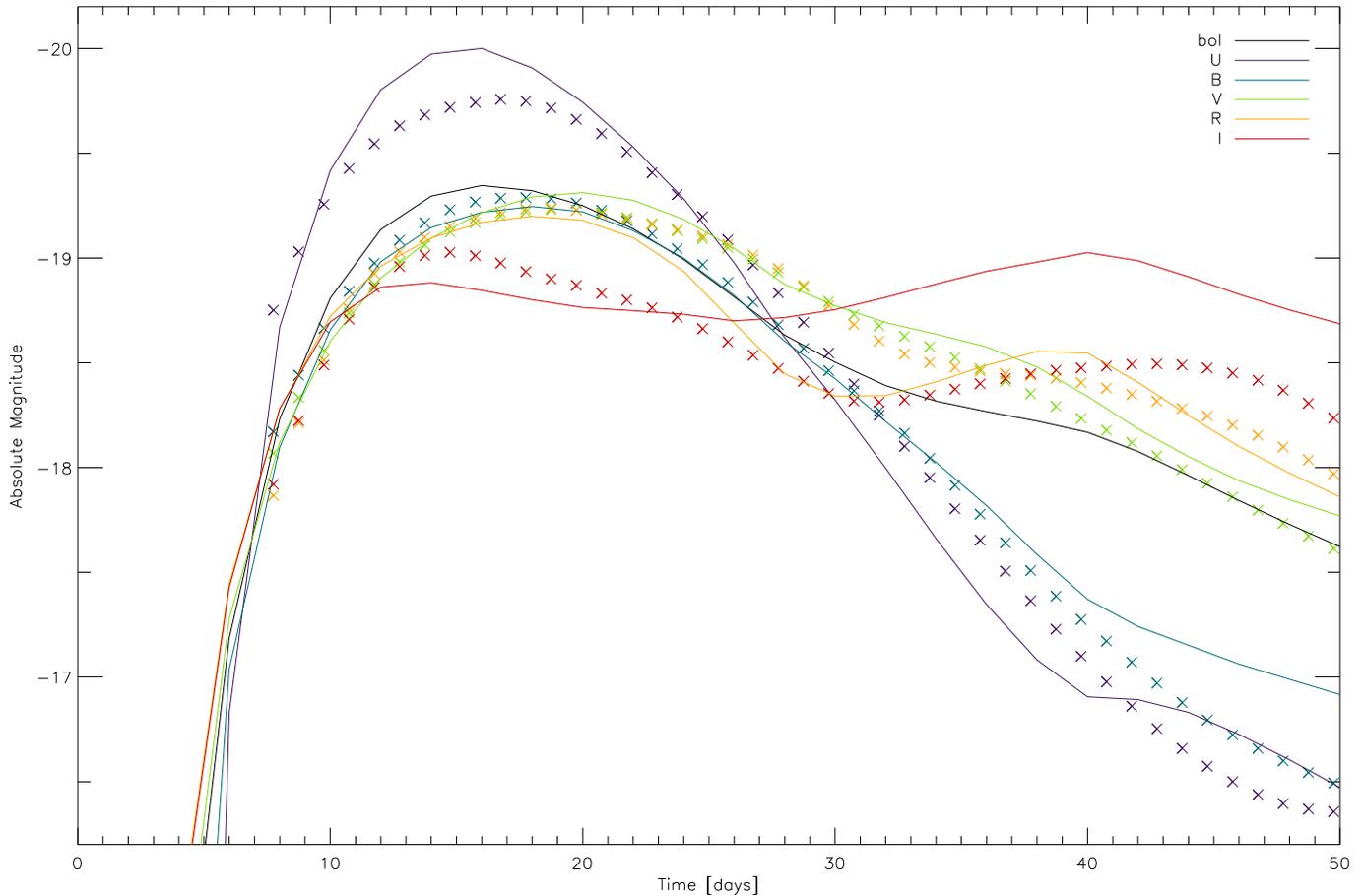}}
 \caption{
 Comparison of the synthetic light curves obtained for the W7 explosion model using the REB method with the MLCS template light curves (see also the close-up in Figure \ref{fig:W7vsMlcsZoom}).
 The templates are based on a large sample of observations, in which each observation is individually corrected for reddening, extinction, etc.
 The one-parameter family of templates plus uncertainties represents the typical observed light curves and the sample variance in them for SNe Ia ranging from high to low peak luminosities.\\
 Direct comparison is made between synthetic and observational absolute magnitudes.
 The maximum B-band magnitudes are aligned in time and the MLCS parameter $\Delta=0.08 \pm 0.03$ is fixed by matching the template and model maximum band magnitudes.
 Thus apart from $t_{\rm max}$ and $\Delta$, which are tightly constrained, there are \emph{no free parameters, no vertical shifts applied}.\\
 The overall resemblance is good, especially given the absence of any free parameters in the fit.
 } \label{fig:W7vsMlcs}
\end{figure*}

\begin{figure}
 \centerline{\includegraphics[width=.5\textwidth]{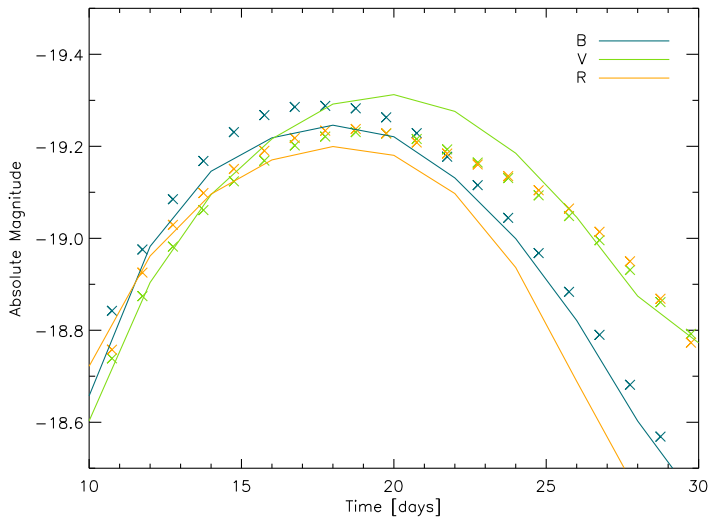}}
 \caption{This close-up of the B, V and R-band light curves of Figure \ref{fig:W7vsMlcs} around maximum brightness shows that the colors of the model light curves (solid lines) reproduce the (observational) MLCS light curves (crosses) reasonably well.
 } \label{fig:W7vsMlcsZoom}
\end{figure}

Light curves and spectra for the W7 explosion model structure obtained using the REB method are shown in Figures \ref{fig:W7vsMlcs} to \ref{fig:W7vsHsiao07} in comparison with observational data.

In Figure \ref{fig:W7vsMlcs} the model UBVRI light curves (we adopt the standard Bessel $[$1990$]$ passband filters) are compared with MLCS light curve templates, as presented in \cite{Jha07}, and a close-up of BVR around maximum brightness showing the color properties is given in Figure \ref{fig:W7vsMlcsZoom}.%
The MLCS model assumes that the SN Ia light curve shape is correlated with the peak luminosity and can be described by the one-parameter function,
\begin{multline} \label{eq:MLCSTemplate}
 {\bf M}_X (\Delta) = {\bf M}_X^0 + {\bf P}_X \Delta + {\bf Q}_X \Delta^2 +\\
   5 \cdot \rm{log}_{10} \left( \frac{H_0}{65 \,\rm{km}/\rm{s}/\rm{Mpc}} \right) \; ,
\end{multline}
where ${\bf M}_X$ is the template magnitude in photometric band $X$ and $H_0$ is the value of the Hubble constant.
${\bf M}_X$ depends only on the intrinsic luminosity-shape parameter $\Delta$, and is a vector in the time domain.
The vectors ${\bf M}_X^0$, ${\bf P}_X$, and ${\bf Q}_X$ are determined by a large sample of observations.
Construction of the template is based on many observations , which provides a way to statistically minimize the uncertainties and the influence of the peculiarities of individual observations and objects.
The templates describe, given the underlying approximation, the best observational knowledge about the typical shape and absolute magnitude of SN Ia light curves and their dependence on peak luminosity.

The absolute observed magnitudes, provided by the templates, can directly be compared with the absolute magnitude obtained from the ballistic phase models.
There are no observational uncertainties like distance modulus or extinction in the templates.
This means that, after aligning horizontally the time of observed maximum brightness to the time of maximum brightness in the model light curves and choosing the maximum template magnitudes to correspond with the model using $\Delta$, there are \emph{no free parameters} that allow scaling of or offsets in the magnitude, either in single bands or in all bands simultaneously.

The obtained value for the luminosity-shape parameter is $\Delta = 0.08 \pm 0.03$ with $H_0 = 72$ km/s/Mpc.
This is a rather central value compared to the values for the observations used in the MLCS training, ranging from -0.40 to 1.38, where $\Delta \lesssim -0.15$ is considered a ``slow decliner'' and $\Delta \gtrsim 0.3$ a ``fast decliner'' (see \cite{Jha07}).

The W7 light curves show good resemblance with the fitted MLCS light curves, meaning that the typical behavior of objects with similar brightness is reproduced by the W7 model using the REB method.
The most notable discrepancy is the late time behavior (after day 25pe) of the I-band; the synthetic flux in the I-band clearly is too large.

\subsection{Spectra}
\begin{figure*}
 \centerline{\includegraphics[width=0.95\textwidth]{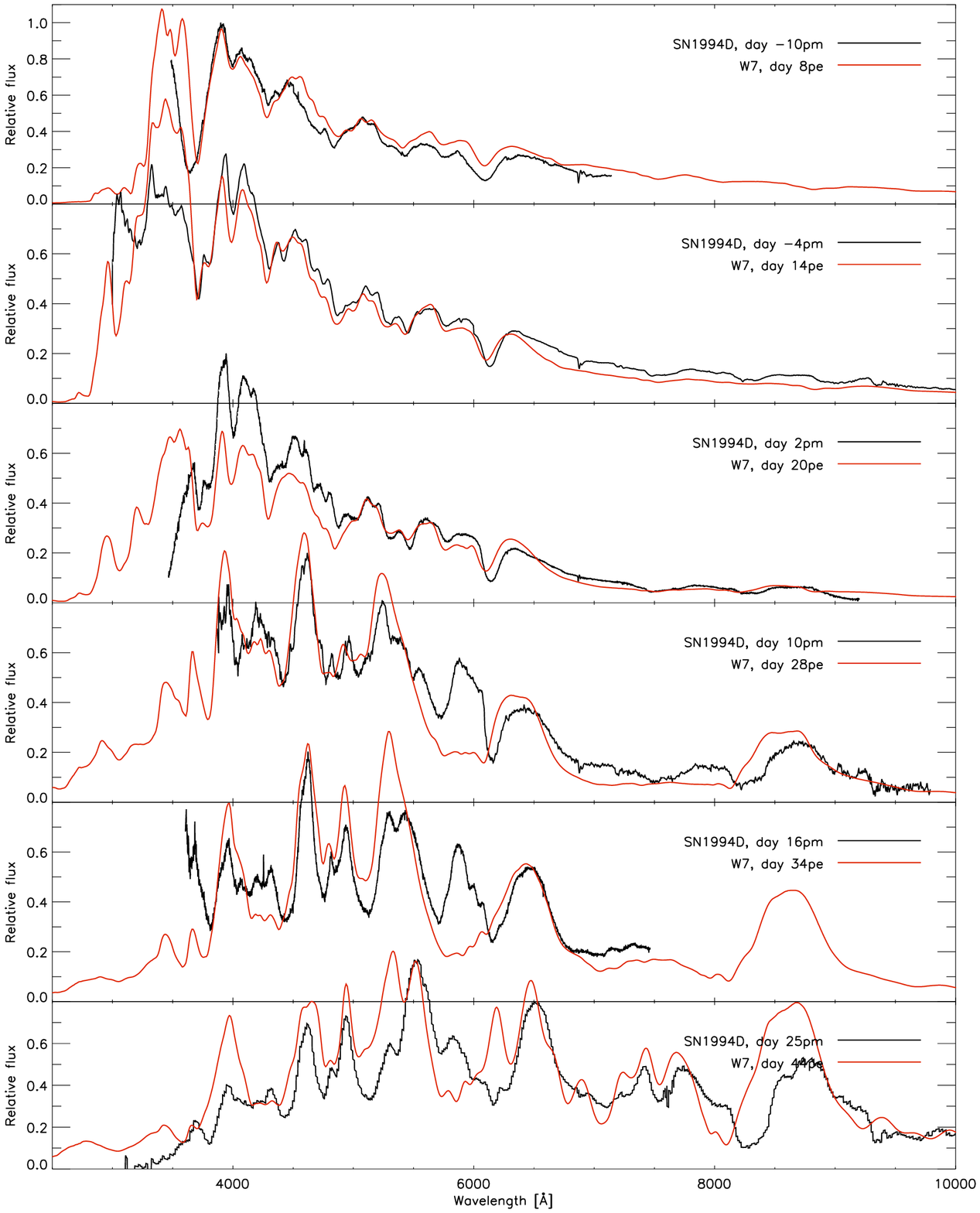}}
 \caption{
 Comparison of the synthetic spectra obtained for the W7 explosion model using the REB method with observed spectra for SN 1994D.\\
 The date of the maximum B-band magnitude in the synthetic light curves [day 18 post-explosion (18pe)], is aligned with day 0 after maximum B-band brightness (0pm).
 A single factor is used to scale \emph{all} observed spectra simultaneously to match the synthetic fluxes.
 Apart from this factor there are \emph{no free parameters} in this comparison.
 The good overall agreement of the fluxes shows that the evolution of the observed luminosity as a function of time is well reproduced using \phx{} and the REB method.
 } \label{fig:W7vs94D}
\end{figure*}

In Figure \ref{fig:W7vs94D}, the model spectra are compared with observed spectra for SN 1994D (source: SUSPECT database).%
This was a nearby event, observed with very good spectroscopic temporal coverage starting as early as 11 days before maximum B-band brightness, is little affected by dust extinction, has a high S/N ratio, and is often compared with W7 model spectra in the literature.
The observed fluxes have together been scaled with a \emph{single} free parameter to account for the distance modulus.
Although not all of the features match well, the overall shapes of the observed spectra are reproduced well by the model spectra.
Also, the luminosities of the spectra correspond well over the whole range in time, from 10 days before to 25 days after maximum B-band brightness.
This indicates that the temporal evolution of the model luminosity accurately reproduces the observed evolution of the luminosity for this benchmark object.

\begin{figure*}
 \centerline{\includegraphics[width=.95\textwidth]{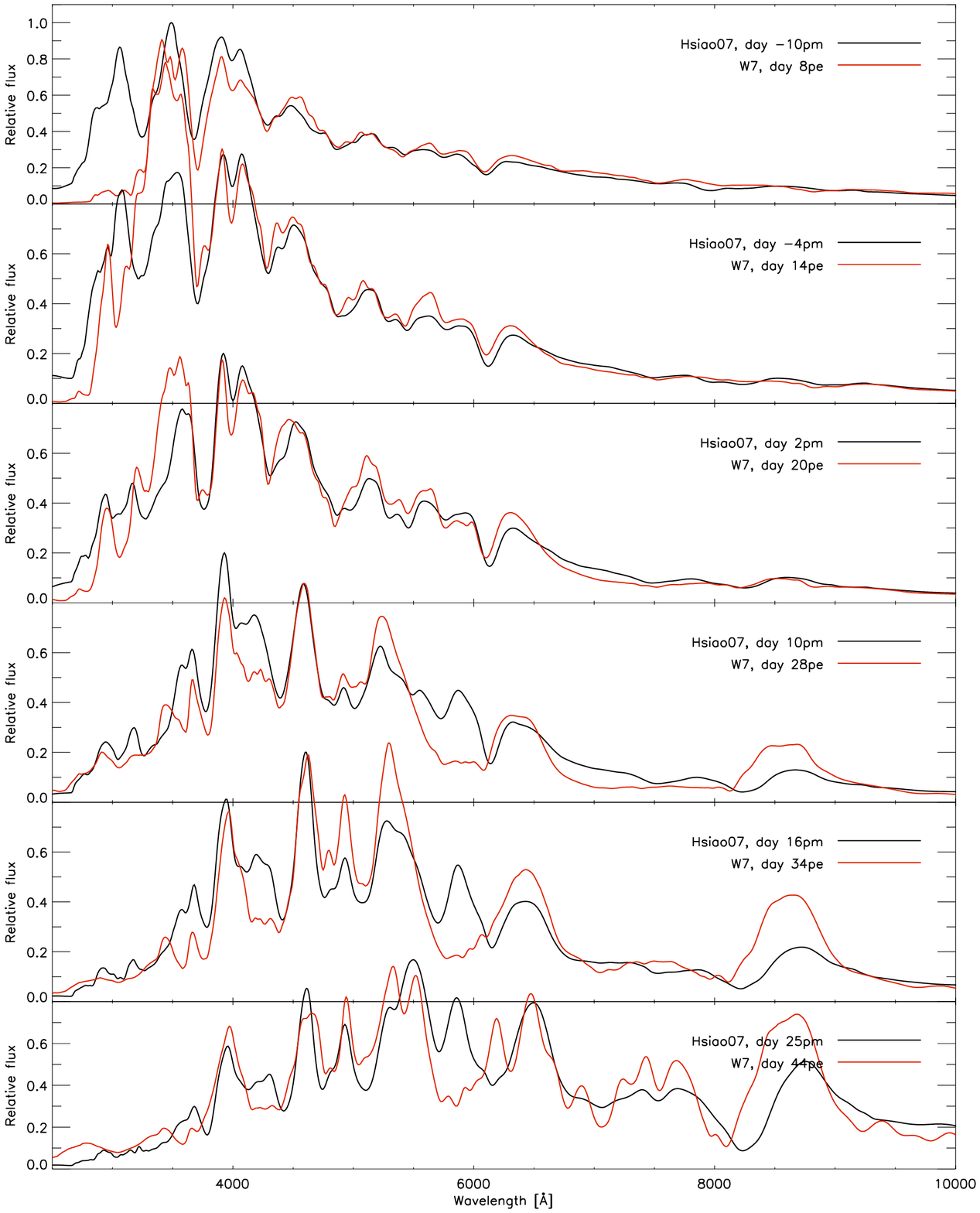}}
 \caption{
 Comparison of the synthetic spectra shown in Figure \ref{fig:W7vs94D} with the spectral templates from \cite{Hsiao07}.
 The templates are averages over a large number of observations of SNe Ia, in order to let characteristic spectral features dominate over the peculiarities of single objects.\\
 A single factor is used to scale \emph{all} of the templates simultaneously to match the synthetic fluxes.
 Apart from that there are \emph{no free parameters} in this comparison.
 The good overall match of the fluxes shows that the evolution of the observed luminosity as a function of time is well reproduced using \phx{} and the REB method.
 } \label{fig:W7vsHsiao07}
\end{figure*}

Single observations (and also single objects) can have peculiar features that may make comparison look better or worse than if other observations would have been chosen.
Therefore, we also compare the synthetic spectra with the spectral templates from \cite{Hsiao07} (see Figure \ref{fig:W7vsHsiao07}).
These templates are averages over many observations, so that characteristic observational features dominate peculiarities.

The synthetic spectra show a good resemblance to the templates both in terms of individual features and of the overall flux.
Narrow features apparent in the synthetic spectra (e.g., between 4500Å and 5000Å in day 14pe) are missing or more ``washed out'' in the templates but are present in the SN 1994D spectra in Figure \ref{fig:W7vs94D}.
In the synthetic spectra the emission feature around 8500Å is too strong on day 28pe and later.

\section{Discussion and Conclusions}
Up until recently, only snapshots in time could be done with Phoenix [see, e.g., \cite{Nugent97}].
Such snapshots are not physically self-consistent because they do not take into account the effects of the prior evolution, including the storage (and later release) of energy in the radiation field.
\cite{Jack11,Jack12} propose a method to overcome these shortcomings but, while the method is based on the GEE and ignores the energy density of the radiation field, it suffers from serious physical and numerical problems as described in sections \ref{sec:ProblemOutline} and \ref{sec:ModifiedRE}.
The REB method proposed in this paper enables the \phx{} radiation transport code -- for the first time -- to solve the time-evolution of Type Ia supernovae in an physically self-consistent and accurate way.

The discrepancies in comparing the synthetic light curves and spectra calculated using \phx{} and the REB method with observations are similar to those found with other codes (see, e.g., \cite{Kasen06} and \cite{Kromer09}).
In particular, the I-band is known to be very sensitive to the Ca II IR triplet (rest wavelength around 8580Å).
A special treatment for the lines of this triplet can significantly improve the fit (see \cite{Kasen06a}), the application of which is not limited to the W7 model (see, e.g., \cite{Woosley07}).
Note, however, that such tweaking is not based on physical arguments.
Furthermore, NLTE effects may play a role as they are known to be important at later times.
Yet, even though the W7 model is known to reproduce many observed properties reasonably well, it is a phenomenological 1D model lacking the detailed physics of many important aspects of the of the explosion phase, and there is no reason to expect it to match the observations perfectly.

We have shown within the context of the W7 model the contributions made by the four physical processes that play a role in the evolution of the energy balance as a function of time.
The contributions from the adiabatic cooling of the radiation field and the ``battery effect'' are large (much larger than the gradient of the radiative flux) and add at early times ($\lesssim$10pe).  They then become smaller than the flux gradient and start to partially cancel each other, but are still significant ($\gtrsim$10pe $\lesssim$60pe).  Finally, they become so small that the flux transports away essentially all of the energy that is deposited locally ($\gtrsim$60pe).
It is only after this time that the energy balance becomes time-independent because it's memory of the past fades away.
Also it is only after about this time that the total amount of energy lost to radiative flux since day 0pe becomes larger than the amount lost to adiabatic cooling through expansion of the radiation field.
Neglecting these effects, as is done in snapshot calculations, will lead to incorrect temperature structures and therefore incorrect spectra.
Furthermore, we have shown that the radiation-field battery term integrates to 0 over time.
This is an important verification of the accuracy of the REB method and our implementation of it, since the energy-balance equation at the heart of the method is in differential form in $t$, not in integral form.

Generalizing the REB method to multi-dimensions is not straightforward.
In multi-dimensions, there is a tangential flux in addition to the radial flux.
Therefore, the target flux is no longer a scalar that can be obtained by radial integration, but a vector.

\acknowledgements
We thank Don Lamb and Eddie Baron for their very helpful comments on the draft of this manuscript and John Dombeck for providing his great IDL color tables.  This research was supported in part by the NSF under grant AST-0909132.

\bibliography{SNsynt}

\end{document}